\documentclass[twocolumn,tightenlines,prd,floatfix,superscriptaddress,nofootinbib]{revtex4}

\usepackage{epsfig} 
\usepackage{amssymb} 
\usepackage{amsmath} 
\usepackage{amsfonts}

\newcommand{\be}{\begin{equation}}
\newcommand{\ee}{\end{equation}}

\begin{document}

\title{Quark mass effects in high energy neutrino nucleon scattering}

\author{Yu Seon Jeong and Mary Hall Reno}
\affiliation{Department of Physics and Astronomy, University of Iowa, Iowa City, IA 52242}

\begin{abstract}
We evaluate the neutrino nucleon charged current cross section at next-to-leading
order in quantum chromodynamic corrections in the variable flavor
number scheme and the fixed flavor number scheme, taking into
account quark masses. The number scheme
dependence is largest at the highest energies considered here,
$10^{12}$ GeV, where the cross sections differ by approximately
13\%. We illustrate the numerical implications of the inconsistent
application of the fixed flavor number scheme.

\end{abstract}

\maketitle

\section{Introduction}

Inelastic scattering experiments with leptons interacting with proton and neutron targets  and hadron-hadron collider results have provided a picture of the structure of the nucleon
over a wide range of momentum transfers. These data combined have led to a parton model picture with parton distribution functions (PDFs) extracted by a number of groups \cite{cteq6,cteq6hq,cteq66,GJR,GJRV,mrs}.
With large underground neutrino telescopes designed to detect neutrinos from astrophysical sources \cite{superk,amanda,icecube,antares}, the PDF inputs to the neutrino-nucleon cross section
are essential ingredients to uncover features of the astrophysical sources. 

Heavy quark masses and their roles in the theory of the
structure functions and the extraction of parton distribution functions at
next-to-leading order
have been explored extensively 
\cite{Gottschalk,vanderBij,Smith,Kramer,Gluck,Kretzer,tung,AOT,ACOT,acotchi,tks,tr,forte,nt,leshouches}.
As neutrino telescope analyses become more refined, it is useful to consider the quark mass effects in the evaluation of neutrino-nucleon
cross sections. Furthermore, in considering ultrahigh energies, 
one can explore the implications of potentially large $\log(Q^2/m_Q^2)$
corrections from non-zero quark masses.

We use the neutrino-nucleon charged current scattering example to discuss two different theoretical approaches to including heavy quark contributions, including next-to-leading order quantum chromodynamic (NLO QCD)
corrections. 
As the incident neutrino energy increases, the average momentum transfer $Q$ also increases, so heavy quarks become effectively ``light'' flavors. We look at the fixed flavor number scheme (FFNS) where the number of light quark flavors is fixed, regardless of the scale of $Q^2$. The Gluck, Reya and Vogt (GRV) \cite{GRV} 3-flavor PDFs are useful for this scheme. Gluck,
Jimenez-Delgado and Reya (GJR) \cite{GJR} have updated these 3-%
flavor PDFs. The variable flavor number scheme (VFNS) allows for the introduction of charm and bottom quarks as constituents of the nucleon as $Q^2$
increases. We use the CTEQ6.6M PDFs
 \cite{cteq66} and a version of the GJR PDFs \cite{GJRV} which are applicable
in the VFNS. The GJR variable flavor PDF set is generated radiatively from 
the 3-flavor set at a factorization scale set by the heavy quark mass
$m_Q$. The GJR variable flavor PDF set is not fit to data beyond the
initial three-flavor fit.

The modified minimal subtraction scheme for NLO corrections (${\overline{\rm MS}}$) with massless quarks is a variable flavor
scheme which neglects the quark mass except effectively with a
step-function threshold factor. A version of the 
Avizas, Collins, Olness and Tung
prescription (ACOT) \cite{AOT,ACOT} is the approach to include
heavy flavor \cite{tks} in the variable flavor number scheme 
that we use here, although there are other approaches to
incorporating the quark mass effects in the generalized mass VFNS \cite{tr,forte,nt,leshouches}.

In Section II, we review the formalism for the neutrino-nucleon charged current cross section at leading order and next-to-leading order in QCD,
and the
ACOT formalism for including quark mass effects is described.
At ultrahigh energies,
the parton distribution functions are probed at very small momentum fractions, in some cases beyond the PDF fits. We discuss our extrapolation to smaller momentum fractions in Section II. 

Our results for the the NLO QCD corrected neutrino-nucleon charged current cross sections are shown in Sec. III. 
The focus in this paper is on the high energy regime, however, we also
show the 100 GeV to $10^4$ GeV energy regime, where one can see effects of flavor scheme, prescription choice and PDF choice. The NLO QCD correction is on the order of 3\% in this energy regime. At ultrahigh energies, quark mass effects (with the exception of the top quark) are negligible so the massless
${\overline{\rm MS}}$ formalism represents the VFNS. The VFNS accounts for a resummation of logarithms $\ln(Q^2/m_Q^2)$ which the FFNS neglects.

As we see numerically in Sec. III,
the FFNS with only three light quarks yields an enhancemnt of the VFNS neutrino-nucleon charged current cross section by more
than 10\%. Already by neutrino energies of $\sim 10^7$ GeV, discrepancies appear in the cross section calculated with and without the
resummation of these logarithms. 
We
summarize our results in Sec. IV, where we also show the numerical implications
of a mismatch in scheme at ultrahigh energies \cite{Basu}.

\section{Neutrino nucleon charged current scattering}

\subsection{Leading-order cross section}

In neutrino-nucleon charged current scattering, one is interested in the
inclusive process. We evaluate the cross section for neutrino scattering with
isoscalar nucleons $N=(n+p)/2$. For definiteness, we consider muon neutrino scattering. For the energies of interest here, we neglect the muon mass. The effect of the tau mass
in $\nu_\tau N$ scattering is on the order of 5\% at $E_\nu=10^3$ GeV \cite{kr}, so the ``low energy'' results reported here are not applicable to tau neutrino charged current scattering.

With the momentum assignments,
\begin{equation}
\nu_\mu(k) + N(p) \rightarrow \mu (k') + X\ ,
\end{equation}
one defines the variables
\begin{eqnarray}
Q^2 &=& -q^2 = -(k-k')^2\\
x &=& \frac{Q^2}{2p\cdot q}\\
y &=& \frac{p\cdot q}{p\cdot k}\ .
\end{eqnarray}
In the massless quark and massless target limit, $x$ is the momentum fraction of the nucleon carried by the struck parton.

The neutrino cross section, in terms of the structure functions $F_1$, $F_2$ and
$F_3$, is
\begin{eqnarray}
\nonumber
\frac{d\sigma}{dx\, dy} &=& \frac{G_F^2 M_N E_\nu}{\pi }\frac{M_W^4}{(Q^2+M_W^2)^2}
\Biggl[ xy^2 F_1 \\ &+ & (1-y - \frac{M_{N}xy}{2E_{\nu}})F_2 
\pm  xy(1-\frac{y}{2})F_3\Biggr]
\label{eq:dsdxdy}
\end{eqnarray}
where $M_W$ is the mass of the $W$ boson and $M_N$ is the mass of the target
nucleon. The upper sign
in Eq. (\ref{eq:dsdxdy}) is for neutrinos, 
the lower sign for anti-neutrino scattering.

At leading order, the quark mass and nucleon mass introduce two types of 
corrections which appear in the structure functions.
In the massless limit, the light cone momentum fraction in the PDFs is the momentum
fraction $x$. Including mass effects, the light cone momentum fraction changes to
\begin{eqnarray}
\xi &=& \eta \frac{Q^2 - m_1^2+m_2^2+\Delta(-Q^2,m_1^2,m_2^2)
}{2Q^2}\\
\eta &=& \frac{2x}{1+\sqrt{1+4x^2M^2/Q^2}} \\ 
\Delta(a,b,c) &=& \sqrt{a^2+b^2+c^2-2(ab+bc+ac)}
\label{eq:xi}
\end{eqnarray}
in terms of initial quark mass $m_1$ and final quark mass $m_2$.
The light cone momentum fraction $\xi$ goes into the quark
distribution function evaluation. We have kept the nucleon mass correction
in $\xi$ (through $\eta$) and in the differential cross section. Below $E_\nu=100$ GeV,
more target mass effects should be included \cite{tmcreview}.

A variant to this approach is to include the kinematic suppression associated with quark masses is through the variable $\chi$ defined by \cite{acotchi}
\begin{equation}
\chi = \eta \Biggl( 1+\frac{(m_1+m_2)^2}{Q^2}\Biggr)\ .
\label{eq:chi}
\end{equation}
For neutral current scattering of a neutrino with a $\bar{c}$, this accounts for the fact that an associated $c$ is also a component of the nucleon,
leading to $\chi_c=\eta(1+4 m_c^2/Q^2)$. Eq. (\ref{eq:chi}) extends this to charged current interactions where final quark and the
remaining sea component quark masses are different. We discuss the numerical difference between using $\xi$ and
$\chi$ in the NLO cross section below.

The second contribution comes
in the mass corrections to the structure functions in terms of
the quark distributions. At leading order, neglecting
corrections proportional to $1+4 x^2 M^2/Q^2$, the mass corrections
are \cite{AOT}
\begin{eqnarray}
\nonumber
F_1 &=&  \sum_{ij} V_{ij}^2\Biggl( \frac{Q^2+m_i^2+m_j^2}{\Delta}\Biggr)\Bigl(
q_i(\xi,\mu^2)+
\bar{q}_j(\xi,\mu^2)\Bigr)\\
\nonumber
F_2 &=& \sum_{ij} 2 x V_{ij}^2 \frac{\Delta}{Q^2}\Bigl(
q_i(\xi,\mu^2)
+ \bar{q}_j(\xi,\mu^2)\Bigr)\\
F_3 &=& \sum_{ij} 2V_{ij}^2 \Bigl(
q_i(\xi,\mu^2)-\bar{q}_j(\xi,\mu^2)\Bigr)
\label{eq:f123lo}
\end{eqnarray}
for an initial quark $q_i$ which converts
to quark $q_j$ or the corresponding antiquark initiated process.  The quantity
$V_{ij}$ is the element of the Cabibbo-Kobayashi-Maskawa (CKM) matrix. We use the
central values quoted by the Particle Data Group \cite{pdg}. For neutrino
scattering, $i=d,\ s,\ b,\ j=\bar{u},\ \bar{c}$ in the variable flavor number
scheme. In the fixed flavor number scheme, $i= d,\ s,\ j=\bar{u}$
for the PDFs, however all flavors are included in the final state sum for the CKM matrix elements. 
The quantity $\mu$ is the factorization scale, which we set to $\mu=Q$ in our
numerical evaluations.

\subsection{Next-to-leading order corrections}

\begin{figure}
\includegraphics[width=3.0in]{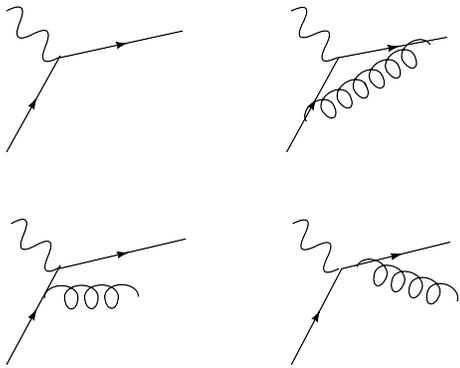}
\caption{The NLO loop correction to the structure functions comes from the interference of the two upper diagrams diagrams. An additional correction comes
from the two lower diagrams.}%
\label{fig:quarks}%
\end{figure}

\begin{figure}
\includegraphics[width=3.0in]{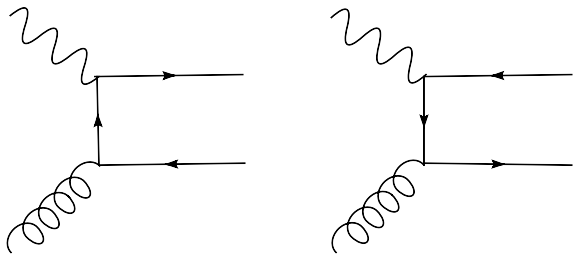}
\caption{Gluon splitting produces a quark-antiquark pair. For the fixed flavor number
scheme with $N_f=3$, this process is the only way to get charm quark and $b$ quark
contributions to the cross section.}%
\label{fig:gluons}%
\end{figure}

At next-to-leading order in QCD, the structure functions $F_i$ have corrections which account for perturbative loop corrections and splitting of quarks, antiquarks and gluons in the 
nucleon. Graphically, the loop corrections and the quark splitting corrections are shown in Fig. \ref{fig:quarks}. In addition to 
the quark initiated contributions shown in this figure, there are also
antiquark initiated diagrams. The gluon contribution to the NLO neutrino-nucleon cross section comes from the graphs in Fig. 
\ref{fig:gluons}.

The final evaluation of the structure functions and ultimately the cross section requires 
a subtraction for factorization \cite{factorization}. The ACOT scheme uses quark masses to regulate collinear divergences. The proof of factorization by Collins in 
Ref. \cite{factorization}
is valid for any scale $Q$ relative to $m_i$. The ACOT prescription, for
example at NLO for the gluon splitting in Fig. \ref{fig:gluons} with $W^* G\rightarrow b\bar{c}$, has subtraction terms proportional to the gluon
splitting coefficient $P_{G\to q}$ times $\ln \mu^2/m_b^2$ and $P_{G\to q}$ times $\ln \mu^2/m_{\bar{c}}^2$.
Generically, for example for the
structure function $F_1$,
\begin{eqnarray}
\nonumber
F_1 &=& \sum_{ij} q_i \otimes \omega^{(0)}_{ij} + \bar{q}_j\otimes \omega^{(0)} _{\bar{i}\bar{j}}\\
\nonumber
 &+& q_i\otimes (\omega^{(1)}_{ij} - \omega_{ij}^{(0)}\frac{\alpha_s}
 {2\pi}P_{i\rightarrow i}\ln\frac{\mu^2}{m_i^2})\\
 \nonumber
 &+& \bar{q}_j\otimes (\omega^{(1)}_{\bar{j}\bar{i}} - \omega_{\bar{j}\bar{i}}^{(0)}\frac{\alpha_s}
 {2\pi}P_{{\bar{j}\rightarrow {\bar{j}}}}\ln\frac{\mu^2}{m_j^2})\\
 \nonumber
  &+& G \otimes\Bigl(\omega_{g\to i \bar{j}}^{(1)} -\omega_{ij}^{(0)}\frac{\alpha_s}{2\pi}
 P_{G\rightarrow i}\ln\frac{\mu^2}{m_i^2}
 \\
 &-& \omega_{\bar{j}\bar{i}}^{(0)}\frac{\alpha_s}{2\pi}
 P_{G\rightarrow \bar{j}}\ln\frac{\mu^2}{m_j^2}
 \Bigr)
 \label{eq:fac}
 \end{eqnarray}
in terms of convolution integrals denoted by the symbol $\otimes$.

For massless quarks in the modified minimal subtraction scheme
$\overline{\rm MS}$, the logs are replaced by $1/\epsilon$ where
dimensional regularization
($d=4-\epsilon$) is used to regulate the infrared and collinear divergences. Even using small quark masses in eq. (11) for the gluon terms ($g\to q_i\bar{q}_j$) and the massless
$\overline{\rm MS}$ for the quark terms (the S-ACOT scheme), the numerical results using massless 
$\overline{\rm MS}$ are reproduced. 
With the ACOT and S-ACOT approaches where the charm and bottom quark masses are explicit, one can use the same formalism
for a range of energy scales as the role of the quark changes from ``heavy'' to ``light'' \cite{kos}. Of course, the up, down and strange quarks are always ``light'' quarks in our evaluation.

We use the ACOT($\chi$) (or S-ACOT($\chi$), as labeled) prescription \cite{acotchi} for the inclusion of all quark masses.
In the ACOT prescription for VFNS evaluations, the minimum $z$ depends on whether or not the PDF in the convolution
is for a gluon or quark/antiquark.
The limits are
\begin{eqnarray}
z_{min} &=& \xi\quad{\rm for\ }{f=q_i,\ \bar{q}_j}\\
\nonumber
z_{min} &=& \chi\equiv\eta \Biggl(\frac{Q^2+(m_i+m_j)^2}{Q^2}\Biggr)\\
&& \quad
{\rm for\ } f=G
\end{eqnarray}
where $\xi$ and $\eta$ are defined in eqs. (6) and (7).
In the subtraction terms, the massless splitting functions are used, however masses are kept in the coefficient functions $\omega$. 
Eq. (\ref{eq:fac}) shows general terms for the splitting functions and
coefficient functions, but we note that at leading order, e.g.,
$P_{i\to i} = P_{\bar{j}\to \bar{j}}$ in the massless quark limit. The ACOT($\chi$)
prescription replaces $\xi\rightarrow \chi$ in both the integration limits and
in the PDF of the leading order term.

The large logarithms associated with the quark mass terms
can cause numerical issues at high energies. We have made a series of
approximations in our evaluation of the variable flavor number scheme
to avoid these numerical problems. We
keep all the masses at low energies, and at the highest energies, we evaluate the
cross section with zero quark masses.

The S-ACOT($\chi$) prescription is used for neutrino energies above $E_\nu=10^6$
GeV. As noted above, the S-ACOT($\chi$) prescription uses the massless $\overline{\rm MS}$ scheme for the quark initiated terms. The subtraction term for the gluon splitting to quarks also has the quark masses set to zero, however, the quark masses are retained in the gluon splitting process \cite{kos}.
For $E_\nu\geq 10^9$ GeV, we use the massless $\overline{\rm MS}$ scheme,
that is, the zero-mass version of the variable flavor number scheme.
These approximations are reliable for a wider range of energies than used
here \cite{os}.

The ACOT scheme and its variants are only one in a class of approaches \cite{tr,forte,nt} 
to including 
quark mass corrections in a general mass, variable flavor number context. One alternative is the Thorne-Roberts (TR) method which uses $Q^2=m_Q^2$ as a transition point for matching fixed flavor structure functions below the heavy quark mass ($m_Q$) to variable flavor structure functions above the transition point\cite{tr}.
Another alternative is the FONLL method applied to DIS \cite{forte}. The various methods differ by terms which are subleading, but which nevertheless may have phenomenological implications. Ref. \cite{nt} considers a rescaling alternative
to $\chi$. A detailed comparison of the heavy flavor
contributions in different generalized mass variable flavor number schemes to heavy flavor contributions to 
deep-inelastic electroproduction structure functions $F_2$ and
$F_L$, with and without $\chi$-scaling, appears in Sec. 22 of Ref. \cite{leshouches}. 

In Sec. III below, we focus on the ACOT($\chi$) scheme and its variants, and a comparison of the high energy neutrino cross section evaluated with this specific generalized mass variable flavor number scheme to a flavor number scheme fixed at low energies. Our primary focus is on how the cross sections differ at high energies, although we show results for energies as low as $E_\nu=100$ GeV. We comment below on the magnitude of the variations in the ACOT schemes (ACOT, S-ACOT and the $\chi$ variants) in
the 100 GeV range as
a rough order of magnitude of the more general scheme dependence. A full comparison of the different generalized mass, variable flavor number schemes is beyond the scope of this paper.

The comparison of a VFNS with the FFNS at high energies is useful to determine
the numerical effects of the VFNS's resummation of $\ln (Q^2/m_Q^2)$.
In the fixed flavor number scheme with flavor number equal to three, 
heavy flavors and light flavors have a separate treatment regardless of the relation between $Q^2$ and characteristic $m_q^2$. The Gluck, Jimenez-Delgado and Reya\cite{GJR} fixed
flavor number scheme PDFs with three light flavors ($u, d, s$) carried in the evolution of the parton distribution functions are used here. 
We denote the FFNS version of the GJR PDFs by
GJRF. Heavy quark contributions ($c,b$) appear only through explicit contributions from gluon splitting. In eq. (\ref{eq:fac}), in the fixed flavor number scheme with $N_f=3$, the heavy quark contributions come in only through $\omega ^{(1)}_{g\rightarrow i\bar{j}}$. There are no subtraction terms corresponding to heavy quarks since there are no heavy quark PDFs.

For variable flavor number scheme evaluations, we use the GJR VFNS version \cite{GJRV} in which the heavy quark constituents
are radiatively generated from the 3 flavor fits to data, 
denoted GJRV. We also use the CTEQ6 \cite{cteq6}
version incorporating heavy quark effects, CTEQ6.6M PDFs \cite{cteq66},
which updates the CTEQ6HQ version incorporating the quark
mass effects through the ACOT prescription \cite{cteq6hq}.

\subsection{Extrapolation to small x}

As we are considering neutrino energies up to the highest energy cosmic rays, $E\sim 10^{12}$ GeV, extrapolations of the PDFs beyond the range of experiments is
required. The PDFs are available numerically for a range of $Q^2$ and for $x_{\rm min}\leq x\leq 1$.
The characteristic $Q^2$ for high energy neutrino-nucleon scattering is $Q^2\sim M_W^2$ since the propagator suppression dominates the evolution of the PDF with increasing $Q^2$ \cite{andreev}. This is well within the range for both the CTEQ and GJR PDFs. The minimum value of $x$ is more constraining at ultrahigh energies. Using
$$ x y (2ME_\nu) = Q^2$$
and $\langle y\rangle\sim 0.2$, $x$ is required below $x_{\rm min}^{\rm CTEQ6.6}=10^{-8}$ for CTEQ6.6M and
below $x_{\rm min}^{\rm GJR}=10^{-9}$ for GJR PDFs. 

We extrapolate the PDFs below $x_{\rm min}$ using a power law
extrapolation \cite{gqrs,lomatch}, where
\begin{eqnarray}
\label{eq:lambda}
\nonumber
x \overline{q}(x,Q^2) = x_{\rm min} \overline{q}(x_{\rm min},Q^2) (x/x_{\rm min})^{-\lambda_{\overline{q}}} \\
x {g}(x,Q^2) = x_{\rm min} {g}(x_{\rm min},Q^2) (x/x_{\rm min})^{-\lambda_{{g}}}\ .
\end{eqnarray}
The antiquark distribution equals the sea quark distribution, and at low $x$, the valence contribution is negligible. For the CTEQ6.6M PDFs, we
have used a $\log(Q)$ dependent form for $\lambda_i$, while for the
GJR sets, we use a constant $\lambda_i$.
The values used for 
$\lambda_i$ are shown in Table \ref{table:lambda}, where for the CTEQ6.6M
set, we show the value for $Q=M_W$. 

\begin{table}[h]
\begin{tabular}{c|c|c|c}
\hline
\hline
	& CTEQ6.6M & GJRV & GJRF \\
\hline
$\lambda_{\overline{u}}$ & 0.276
&0.255 &0.260		\\
\hline
$\lambda_{\overline{d}}$ & 0.276
&0.255 &0.260		\\
\hline
$\lambda_{\overline{s}}$ & 0.276 
&0.255 &0.260 		\\
\hline
$\lambda_{\overline{c}}$ & 0.277
&0.257 & - 		\\
\hline
$\lambda_{\overline{b}}$ & $0.284 $
&0.264 & - 		\\
\hline
$\lambda_{g}$ & 0.292
&0.267 & 0.273							\\
 \hline
$x_{\rm min}$ & $10^{-8}$  & $10^{-9} $ & $10^{-9} $ \\
\hline
\hline
\end{tabular}
\caption{The parameters that appear in the small $x$ extrapolations of eq. (\ref{eq:lambda}). For the CTEQ6.6M set, we use a $\log(Q)$ dependent
form. Here we show the value at $Q=M_W$.}
\label{table:lambda}
\end{table}

Extrapolations using functional forms other than power laws have been suggested by a number of authors \cite{saturation,ranges}.
The typical range of predictions for $E_\nu=10^{12}$ GeV is on the order of a factor of $0.5 - 2$ times the cross sections
reported here.

\section{Results}


In our evaluation of the quark mass effect on the $\nu N$ cross section at NLO, we restrict our attention to the
CTEQ6.6M and GJR PDFs.
We set the heavy quark masses to $m_{c}$= 1.3 GeV, and $m_{b}$= 4.5 GeV for CTEQ6.6M,
and to $m_{c}$= 1.3 GeV, and $m_{b}$= 4.2 GeV for GJR as indicated in each PDF. 
As noted above, we use the ACOT($\chi$) scheme for our evaluations in the variable flavor number scheme, with transitions to S-ACOT($\chi$) and 
$\overline{\rm MS}$ as the neutrino energy increases.

\begin{figure}[h]%
\includegraphics[width=3.0in]{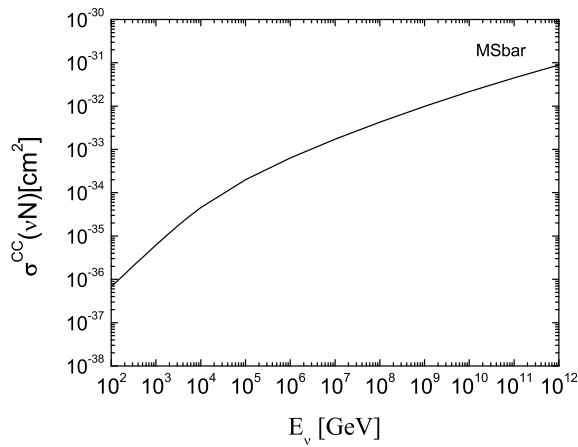}
\caption{The $\nu N$ cross section for the charged current process as a function 
of the incident neutrino energy. The cross section  is evaluated using the massless $\overline{\rm MS}$ scheme
with the CTEQ6.6M PDFs.}%
\label{fig:sig}%
\end{figure}

In Fig. \ref{fig:sig}, we show the $\overline{\rm MS}$ neutrino-nucleon
charged current cross section as a function of the incident neutrino energy.
This sets the scale of the cross section. The CTEQ6.6M PDFs give a cross section, using the
massless $\overline{\rm MS}$ scheme, that are within $ 2\%$ of the
standard CTEQ6 result, using the $\overline{\rm MS}$ scheme at
$E_\nu=100$ GeV, and less than 0.5\% different at $E_\nu=10^{10}$ GeV.
In subsequent figures, we show only ratios.

\begin{figure}[h]%
\includegraphics[width=3.0in]{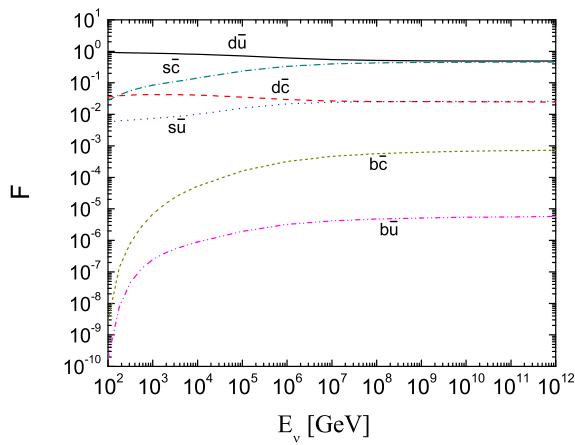}
\caption{The ratio of the separate flavor contributions 
to the neutrino-nucleon
charged current cross section, evaluated using the CTEQ6.6M PDFs in 
the VFNS.}%
\label{fig:F}%
\end{figure}

To get an idea of the relative importance of the flavor components, in
Fig. \ref{fig:F} we show the ratio of the cross section for each sub-process
in the neutrino nucleon charged-current cross section. At NLO, it
is not possible to separate, e.g., the $d$ and $\bar{u}$ contributions since
the gluon splitting diagrams (Fig. \ref{fig:gluons}) are added at the 
amplitude level. The figure shows that the $d\bar{u}$ contribution dominates
until the charm mass corrections and valence contributions are neglible.
Then, the $s\bar{c}$ contribution is nearly equal to the $d\bar{u}$
contribution. At high energies,
the $s\bar{u}$ and $d\bar{c}$ terms are also nearly equal. At low energies, the
valence $d$ component more than compensates for the mass suppressed  charm quark production.
Contributions involving the $b$ quark are at most at the level of
0.1\%.

\begin{figure}[h]%
\includegraphics[width=3.0in]{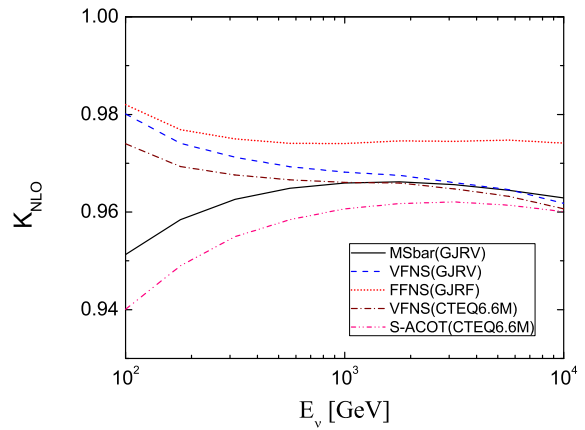}
\caption{K$_{\rm NLO}$-factor, the ratio of the $\sigma_{NLO}$ to $\sigma_{LO}$ (with quark masses) 
for the VFNS and FFNS. The LO cross section is evaluated using the same PDFs as the NLO cross section and the scaling variable $\chi$ is used throughout. 
For VFNS, all quark masses are kept while for the
$\overline{\rm MS}$ curve, quark masses are set to zero.}%
\label{fig:K}%
\end{figure}

Fig. \ref{fig:K} shows K$_{\rm NLO}$-factor, which is the ratio of NLO cross section to LO cross section,
for the incident neutrino energy between $10^2$ GeV and $10^4$ GeV.
In each ratio, the LO cross section is evaluated using the same PDF as the NLO cross section, namely the NLO PDF set, to exhibit the size of the partonic cross section correction. We also use $\chi$ as the scaling variable for all but
the massless $\overline{\rm MS}$ result.
A comparision of the different schemes  and PDFs shows that
at 100 GeV, they differ by as much as $\sim 4\%$.
The K$_{\rm NLO}$-factor of the GJRV VFNS with all masses included
are very close to the massless $\overline{\rm MS}$ results
for $E_\nu>10^3$ GeV,
where the quark masses have little impact. Below this energy,
quark mass effects suppress some of the QCD corrections.
The CTEQ6.6M results are intermediate between the massless
$\overline{\rm MS}$ GJRV and GJRV massive
VFNS results below $\sim 1$ TeV. 

The K$_{\rm NLO}$-factor for the FFNS is only a little higher than the VFNS at $E_\nu =10^2$ GeV,  
but the difference in the K$_{\rm NLO}$-factor increases to about 1.5\% relative to the GJRV K$_{\rm NLO}$-factor at $10^4$ GeV. 
Since the K$_{\rm NLO}$-factor is a ratio, Fig. \ref{fig:K} does not illustrate the fact that the 3 flavor NLO charged
current cross section using GJRF is about 1\% lower than the GJRV cross section. The GJRF ``LO'' cross section is lower than the GJRV ``LO'' cross section
by $\sim 2.5\%$ at $E_\nu=10^4$ GeV (using the NLO PDFs).

Fig. \ref{fig:K} also shows the difference in $K_{NLO}$ between the
ACOT($\chi$) and the S-ACOT($\chi$) prescriptions using the
CTEQ6.6M PDFs at $E_\nu=100$ GeV, amounting to more than 
3\%. 
A similar positive offset of 3\% is seen when one compares the cross section at NLO
using the ACOT or S-ACOT scaling variable $\xi$  rather than the
variable $\chi$. At 100 GeV, this would indicate that the specific variable flavor scheme for incorporating
the quark mass and the scaling variable are more important than the fixed flavor/variable flavor choice.
At higher energies (above 1 TeV), ACOT and S-ACOT NLO cross sections differ by less than 1\%, with the difference between using $\xi$ and $\chi$ for the quark terms less than $1.5\%$. At $E_\nu = 10^4$ GeV, the NLO cross sections for the 
CTEQ6.6M PDF are essentially identical for ACOT, S-ACOT, ACOT($\chi$) and
S-ACOT($\chi$). The differences between
generalized mass, variable flavor number schemes and scaling variable choices
continue to be areas for further work \cite{leshouches}.

\begin{figure}[h]%
\includegraphics[width=3.0in]{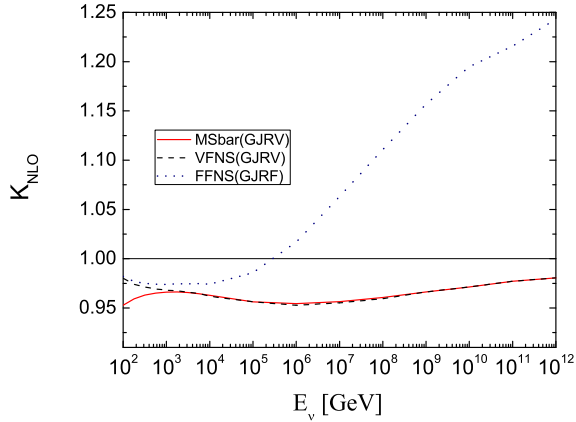}
\caption{K$_{\rm NLO}$-factor: the ratio of the neutrino-nucleon charged current $\sigma_{NLO}$ to $\sigma_{LO}$ for
$E_\nu=10^2-10^{12}$ GeV. The LO cross section is evaluated using the same PDFs as the NLO cross section. The VFNS cross section is evaluated using the
ACOT($\chi$) or S-ACOT($\chi$) scheme for $E_\nu\leq 10^9$ GeV.}%
\label{fig:Kext}%
\end{figure}

In Fig. \ref{fig:Kext}, we show the K$_{\rm NLO}$-factor for the full energy range for
the GJR PDFs. As shown in Fig. \ref{fig:K}, the K$_{\rm NLO}$-factor for the VFNS and $\overline{\rm MS}$ is essentially identical above $E_\nu\sim 10^3$ GeV.
Given the broad energy range of concurrence with VFNS with masses and the massless
$\overline{\rm MS}$, we use the massless $ \overline{\rm MS}$ cross section
for the VFNS above $E_\nu=10^9$ GeV to avoid numerical errors associated with
the subtraction terms. This ``patching'' is used for the remaining figures.

\begin{figure}[h]%
\includegraphics[width=3.0in]{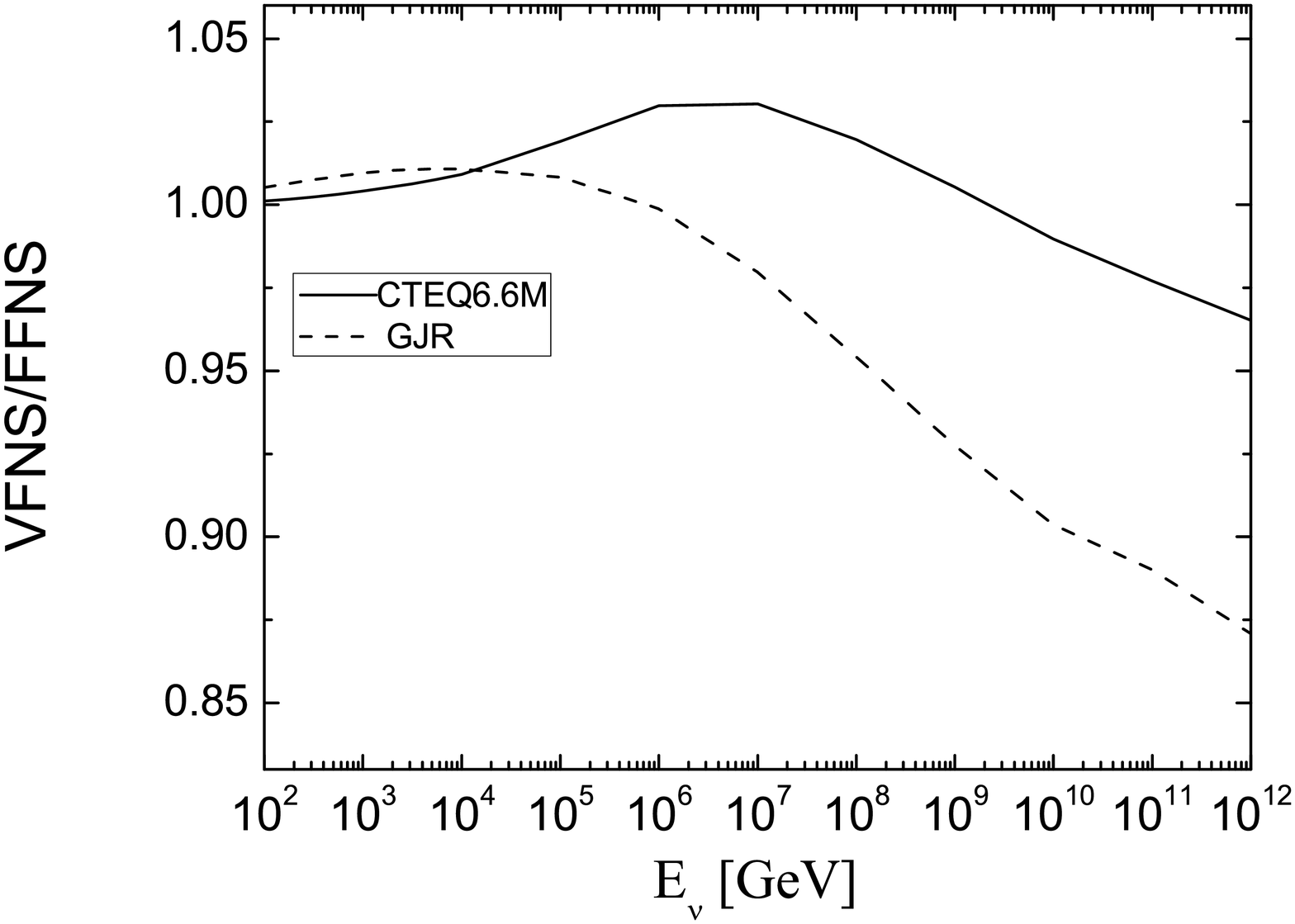}
\includegraphics[width=3.0in]{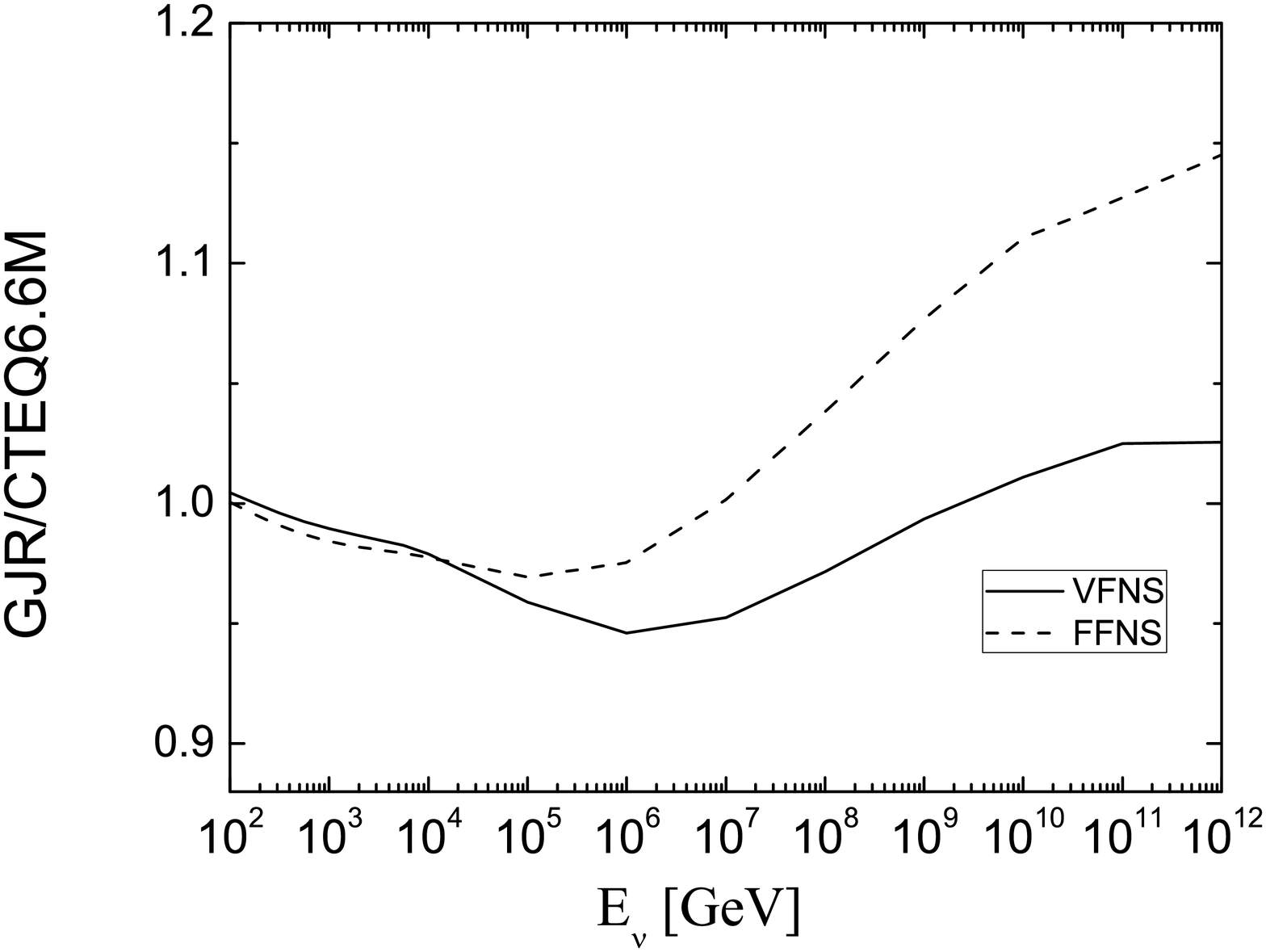}
\caption{(a) The ratio of the NLO charged current
$\sigma_{VFNS}$ to $\sigma_{FFNS}$ for CTEQ6.6M and GJR PDFs.
(b) Comparison of GJR and CTEQ6.6M PDFs for the NLO charged current
$\sigma_{VFNS}$ and $\sigma_{FFNS}$.}
\label{fig:ratio}%
\end{figure}

The K$_{\rm NLO}$-factor for the FFNS starts to deviate from the VFNS K$_{\rm NLO}$-factor at about
$E_\nu\sim 10^4$ GeV, with significant deviations by $E_\nu=10^6-10^7$ GeV
where, in Fig. \ref{fig:F},
the charm quark contribution is effectively ``massless.'' For evaluations at the level of less than 5\% error in the neutrino nucleon cross section, this is the energy range above which the VFNS should be used. This
is shown graphically in Fig. \ref{fig:ratio}.

In Fig. \ref{fig:ratio}(a) we show the ratio of VFNS NLO cross section to FFNS result for different PDFs. The GJRV set is used for the VFNS
and the GJRF set for the FFNS result for the dashed GJR curve.
The CTEQ6.6M set includes 5 quark consituents, so
the ``FFNS'' in this figure for CTEQ6.6M simply omits contributions for $c$ and $b$ quarks and antiquarks, even though
they are consituents of the nucleon in this set.
The GJR sets are a better pair of PDFs to compare, since the two sets are designed to accommodate different flavor numbers.

While the ratio of VFNS to FFNS cross sections in Fig.
\ref{fig:ratio}(a) are stable for energies below $\sim 10^6$ GeV for the GJR PDFs, the
ratio decreases as the incident energy is increased higher, and there is about 13$\%$ discrepancy 
at $10^{12}$ GeV.
For CTEQ6.6M PDFs, they have almost the same value up to $10^{5}$ GeV, and at higher energies their ratio varies
depending on the energy. Their maximum difference is at highest energy, which is about 3$\%$. 

We also compared the cross sections of the VFNS and FFNS for GJR and CTEQ6.6M PDFs. 
As shown in Fig. \ref{fig:ratio}(b), the  
FFNS evaluations using the CTEQ6.6M and GJRF PDFs have very close values up to $E_\nu \sim 10^{6}$ GeV.
At higher energies, however, the FFNS cross section for GJR PDFs grows as the energy increases, 
and it makes difference about 15$\%$ with the result for CTEQ6.6M. 
The more reliable VFNS results for GJRV and
CTEQ6.6M differ by about 5\%  at $E_\nu \sim 10^6
$ GeV, and differ by less than 3\% at $10^{12}$ GeV. 
We note that direct
measurements of deep-inelastic scattering have been done only to an
equivalent neutrino energy of $E_\nu \sim 5\times 10^4$ GeV \cite{hera}.
Up to this energy, the FFNS and VFNS ratios are essentially unity.

\section{Conclusions}

\begin{figure}[t]%
\vskip 0.2in
\includegraphics[width=3.0in]{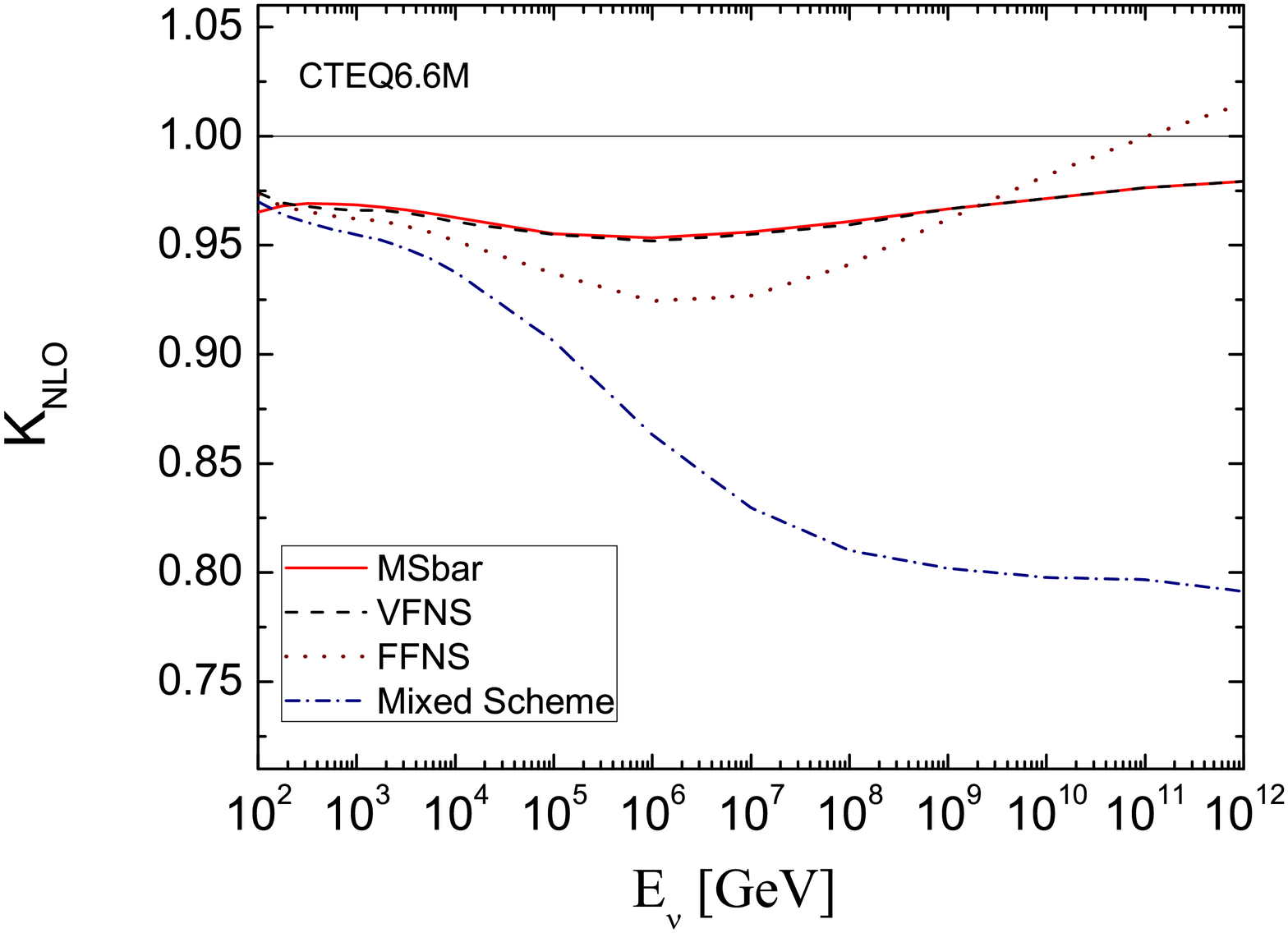}
\caption{K-$_{\rm NLO}$-factor: the ratio of the neutrino-nucleon charged current $\sigma_{NLO}$ to $\sigma_{LO}$ for
$E_\nu=10^2-10^{12}$ GeV. The CTEQ6.6M PDFs are used. For the FFNS
NLO results, the $c$ and $b$ PDFs are set to zero, but all five flavors
of CTEQ6.6M are used for the LO cross section used for all the curves
in the figure.
The ``mixed scheme'' ratio uses 3 flavors of quark PDFs but makes a subtraction for 5 flavors from the gluon fusion term.}%
\label{fig:KextCTEQ6HQ}%
\end{figure}

A concurrence between number scheme, PDF set and related subtraction
terms has been emphasized in the literature, e.g., in Ref. \cite{os,leshouches}.
Mismatches in application can lead to errors on the order of 20\%.
To exhibit one such mismatch, we show the K$_{\rm NLO}$-factor for the
CTEQ6.6M PDFs in Fig. 
\ref{fig:KextCTEQ6HQ}. The dot-dashed line, labeled ``mixed scheme,''
shows the ratio of the NLO cross section to the LO cross section, where
the NLO evaluation has only $d,\ s$ and $\bar{u}$ light quarks but
with the full 5-flavor $\overline{\rm MS}$ gluon subtraction correction.
A similar K$_{\rm NLO}$-factor appears in Ref. \cite{Basu}, in which the 3-flavor
GRV PDFs are used. 

At energies below $E_\nu\sim 10^6$ GeV, the ratio of the VFNS
and FFNS cross sections using the GJR PDFs is $\sim 1$, but at higher 
energies, the ratio drops. The almost $13\%$ discrepancy between the two cross
sections can be attributed, at least in part, to the summation of large 
$\log(Q^2/m_Q^2)$ in the VFNS PDFs. This is a quantitative example of the statement that a three flavor calculation of structure functions overestimates the ``true'' structure functions when more flavors are active \cite{os}.

The GJRV PDFs are not fit to data, but instead generated radiatively
from the 3-flavor fit. The CTEQ6.6M set is fit including mass effects,
so they should be considered more reliable at high energies, especially
where the charm quark contribution is more important. Even so, the discrepancy between the GJRV and CTEQ6.6M VFNS evaluations agree well at the highest energies.

Cooper-Sarkar and Sarkar (CSS) have evaluated the NLO neutrino nucleon cross sections using an independent fit to the data at NLO \cite{css}. 
Our $\nu N$ charged current cross sections using CTEQ6.6M are bigger
than CSS's cross sections by about 8-18\% for 
$s=2M_NE_\nu = 10^8-10^{12}$ GeV$^2$, generally within their estimates
of PDF uncertainty. The GJRV cross section at $s=10^{12}$ GeV$^2$ differs
from CSS by about 20\%.

Uncertainties at the level of a few or a few tens of percent at
$E_\nu=10^{12}$ GeV rely on perturbative QCD and
DGLAP evolution being applicable to very small $x$ values for
$Q^2\sim M_W^2$. At lower values of $Q$, gluon recombination and
saturation effects are important \cite{saturation}, however, at
$Q=M_W$, it is not clear that saturation should be important for the 
total cross section \cite{vogelsang}. As noted in the introduction,
there are a range of predictions that do not rely on DGLAP evolved
PDFs \cite{ranges}. On a short time scale, one looks forward to further
information as the LHC analyses yield PDFs from data in new ranges of $x$ for $Q\sim M_W$, as a start to the experimental probe of PDFs 
and structure functions required
for the ultrahigh energy neutrino cross sections.

\noindent
{\bf Note added:}

Since our submission of this paper, Gluck, Jimenez-Delgado and Reya, in Ref. 
\cite{gjrreply} have emphasized that the K-factor is traditionally defined as the
ratio between NLO and LO (with LO partonic cross sections {\it and} LO PDFs). 
To avoid confusion, we have relabeled our ratio from K in
the original version of this paper to K$_{\rm NLO}$. We confirm the results of Ref. \cite{gjrreply} that the K-factor as traditionally defined
does decline to about $\sim 0.6$ for the VFNS (GJRV) as compared to $\sim 0.8$ 
for the FFNS (GJRF). The authors of Ref. \cite{gjrreply} also comment on the importance of the
$b-\bar{t}$ contribution at ultrahigh energies. Our conclusions about the ratio of VFNS/FFNS at NLO
with the GJR PDFs do not change with the inclusion of the top quark contribution. With the
$b-\bar{t}$ term, the CTEQ6.6M charged current cross section is larger than the CSS cross sections by
13-31\% for $s=10^8-10^{12}$ GeV$^2$.

\begin{acknowledgments}
This research was supported by 
US Department of Energy 
contract DE-FG02-91ER40664. We thank Fred Olness for providing his program to evaluate the structure functions and for useful discussions.
\end{acknowledgments}



\end{document}